\documentclass[11pt,eqs]{article}
\usepackage{latexsym}

\makeatletter
\newcommand\xleftrightarrow[2][]{%
  \ext@arrow 9999{\longleftrightarrowfill@}{#1}{#2}}
\newcommand\longleftrightarrowfill@{%
  \arrowfill@\leftarrow\relbar\rightarrow}
\makeatother

\title{
Noncommutativity and nonassociativity of closed bosonic string on T-dual toroidal backgrounds
\thanks{Work supported in part by
the Serbian Ministry of Education, Science and Technological Development, under contract No. 171031. I also want to thank to Prof. Dr. Branislav Sazdovi\'c and Dr. Ljubica Davidovi\'c from Institute of Physics Belgrade
for useful discussions.}}
\author{B. Nikoli\'c\thanks{email: bnikolic@ipb.ac.rs}\\{\it Institute of Physics Belgrade, University of Belgrade, Pregrevica 118, Serbia} \\ D. Obri\'c\thanks{email: dobric@ipb.ac.rs}\\{\it Faculty of Physics, 
University of Belgrade, Studentski trg 12, Belgrade, Serbia}
}

\begin{document}

\maketitle
\begin{abstract}
In this article we consider closed bosonic string in the presence of constant metric and Kalb-Ramond field with one non-zero component, $B_{xy}=Hz$, where field strength $H$ 
is infinitesimal. Using Buscher
T-duality procedure we dualize along $x$ and $y$ directions and using generalized T-duality procedure along $z$ direction imposing trivial winding conditions. After first two
T-dualizations we obtain $Q$ flux theory which is just locally well defined, while after all three T-dualizations we obtain nonlocal $R$ flux theory. Origin of non-locality
is variable $\Delta V$ defined as line integral, which appears as an argument of the background fields.
Rewriting T-dual transformation laws in the canonical form and using standard Poisson algebra, we obtained that $Q$ flux theory is commutative one and the $R$ flux
theory is noncommutative and nonassociative one. Consequently, there is a correlation between non-locality and closed string noncommutativity and nonassociativity.
\end{abstract}

\section{Introduction}
\setcounter{equation}{0}

Coordinate noncommutativity means that there exists minimal possible length, which imposes natural UV cutoff.
Idea of coordinate noncommutativity is very old. Heisenberg suggested coordinate noncommutativity to solve the problem of the occurrence of infinite quantities before renormalization procedure was developed and accepted. 
The first scientific paper considering this subject appeared 1947 \cite{snyder} where construction of discrete
Lorentz invariant space-time is presented. Later in the period of 1980s A. Connes developed noncommutative geometry as a generalization of the standard commutative geometry \cite{connes}.

Noncommutativity became again interesting for particle physicists when the paper \cite{SW} appeared. In this article it is shown using propagators that open string endpoints in the presence of the constant metric and Kalb-Ramond field
become noncommutative. D-brane on which the string endpoints are forced to move becomes noncommutative manifold. After this article many articles \cite{gbnc} appeared addressing the same subject but using different approaches - Fourier expansion,
canonical methods, solving of boundary conditions etc. 

In the last two articles of \cite{gbnc} the method of solving of boundary conditions is presented. The basic idea is that open string boundary condition is treated as
canonical constraint. Investigating the consistency of the canonical constraint we obtained the $\sigma$ dependent form of the boundary condition. Further, we can proceed twofold: to introduce Dirac brackets or solve the constraint.
Solving the constraint, we obtained the initial coordinate as a linear combination of the effective coordinate and momenta. Consequently, initial coordinates are noncommutative and the main contribution to
noncomutativity parameter comes from Kalb-Ramond field as it was expected.

Following the result of the article \cite{mebo} it can be proven that gauge fields "live" at the open string endpoints. Consequently, many interesting papers 
concerning non-commutative Yang-Mills theories and their renormalisability appeared \cite{ncym}.
In the papers \cite{expnc} cross sections for some decays, allowed in noncommutative Yang-Mills theories and forbidden in commutative ones, are calculated, which offers a possibility of the experimental check of the 
noncommutativity idea and further, indirectly, idea of strings.

It is obvious that closed bosonic string in the presence of constant background fields remains commutative. There are no boundaries and, consequently, boundary conditions constraining string dynamics. In the case of open 
string we obtained initial coordinate in the form of linear combination of effective coordinates and momenta using boundary condition. That is achieved in the closed string case \cite{Lust} using T-duality procedure and 
coordinate dependent background.

T-duality as a fundamental feature of string theory \cite{S,B,RV,GPR,AABL,nasnpb,englezi}, unexperienced by point particle,  makes that  there is no physical
difference between string theory compactified on a circle of radius $R$ and circle of radius $1/R$.
Buscher T-dualization procedure \cite{B} represents a mathematical frame in which T-dualization is realized.
If the background fields do not depend on some coordinates then those coordinates are isometry directions. Consequently, that symmetry can be localized replacing 
ordinary world-sheet derivatives $\partial_\pm$ by covariant ones $D_\pm x^\mu=\partial_\pm x^\mu + v^\mu_\pm$, where $v^\mu_\pm$ are gauge fields. In order to make T-dual theory has 
the same number of degrees of freedom, the new term with Lagrange multipliers is added to the action which forces the gauge fields to be unphysical degrees of freedom.
Because of the shift symmetry, using gauge freedom we fix initial coordinates.
Variation of this gauge fixed action with respect to the Lagrange
multipliers produces initial action and with respect to the gauge fields produces T-dual action.

Standard Buscher T-dualization was applied in closed string case in the papers \cite{Lust,ALLP,ALLP2,L,nongeo1}. In Ref.\cite{ALLP} authors consider 3-torus in the presence of constant metric and Kalb-Ramond field with one nonzero 
component $B_{xy}=Hz$, where field strength $H$ is infinitesimal. They systematically apply Buscher procedure
and, after two T-dualizations along isometry directions, obtain theory with $Q$ flux which is noncommutative. In the calculations they used nontrivial boundary conditions (winding condition). The result is that
T-dual closed string coordinates are noncommutative for the same values of parameters $\sigma=\bar\sigma$ with noncommutativity parameter proportional to field strength $H$ and $N_3$, 
winding number for $z$ coordinate.

But, except this standard Buscher procedure, there is a generalized Buscher procedure dealing with background fields depending on all coordinates.
The generalized procedure was applied to the case of bosonic string moving in the weakly curved background \cite{DS1,DNS2,DNS} and in the case where metric is quadratic in coordinates and Kalb-Ramond field
is linear function of coordinates \cite{DNS3}. The generalized procedure enables us to make T-dualization in mentioned cases along arbitrary subset of coordinates.  

Double space is one picturesque framework for representation of T-duality. Double space is introduced two to three decades ago \cite{Duff,AAT1,AAT2,WS1,WS2}. It is spanned by double coordinates
$Z^M=(x^\mu,y_\mu)$ $(\mu=0,1,2,\dots,D-1)$, where $x^\mu$ are the coordinates of the initial theory and $y_\mu$ are T-dual coordinates. In this space T-dualization is represented as $O(d,d)$ transformation
\cite{Hull,Hull2,berman,negeom,hohmz}. Permutation of the appropriate subsets of the initial and T-dual coordinates is interpreted as partial T-dualization \cite{sazdam,sazda} expanding Duff's idea \cite{Duff}. The newly invented intrinsic noncommutativity
\cite{minic} is related to double space. Intrinsic noncommutativity exists in the constant background case because it is considered within double space framework.

In this article we will deal with closed bosonic string propagating in the constant metric and linear dependent Kalb-Ramond field with $B_{xy}=Hz$, the same background as in \cite{ALLP}. This configuration is known in
literature as torus with $H$-flux. As in the Ref.\cite{ALLP} we will use approximation of diluted flux,
which means that in all calculations we keep constant and linear terms in infinitesimal field strength $H$. Transformation laws, relations which connect initial and T-dual variables, we will write in canonical form expressing 
initial momenta in terms of the T-dual coordinates. Unlike Ref.\cite{ALLP}, except T-dualization along 
two isometry directions, we will make one step more and T-dualize along $z$ coordinate using generalized T-dualization procedure. During dualization procedure we will use trivial boundary (winding) conditions.

Transformation laws in canonical form enable us to express sigma derivative of the T-dual coordinate as a linear combination of the initial momenta and coordinates. Because initial theory is geometrical locally and globally,
its coordinates and canonically conjugated momenta satisfy standard Poisson algebra. This fact means that we can calculate the Poisson brackets of the T-dual coordinates using technical instruction given in 
subsection 4.1.

After T-dualizations along isometry directions (along $x$ and $y$) we obtain the same background as in Ref.\cite{ALLP} but, obtained $Q$ flux theory, which is still locally well defined, is commutative. This is
a consequence of the imposed trivial winding conditions.
Having in mind the generalized T-duality procedure \cite{DS1, DNS2, DNS3}, T-dualization along $z$ coordinate produces $R$ flux nonlocal theory because it depends on the variable $\Delta V$ which is defined 
as line integral. Calculating Poisson brackets of the T-dual coordinates we obtain two nonzero Poisson brackets and show that there is a correlation between non-locality and closed string noncommutativity.

The form of noncommutativity is such that it exists when arguments of the coordinates are different, $\sigma\neq \bar\sigma$. That is another difference with respect to the result of Ref.\cite{ALLP} but there is no
contradiction because the origins of noncommutativity are different. In this article non-locality is related with noncommutativity of $R$ flux theory under trivial winding conditions while in Ref.\cite{ALLP} it is about noncommutativity 
of $Q$ flux theory under nontrivial winding conditions.

From the noncommutativity relations it follows that Jacobi identity is broken i.e. nonassociativity occurs. Nonassociativity parameter, $R$ flux, is proportional to the field strength $H$. Using generalized
T-duality \cite{DS1,DNS2,DNS3} we obtain the concrete form of nonassociativity from string dynamics. Similar as noncommutativity, discovery of nonassociativity pushes the scientist to explore the effects of nonassociativity
in the field of renormalisability of $\phi^4$ theory \cite{MTMY} as well as formulation of nonassociative gravity \cite{nagravity}. 

At the end we add an appendix containing some conventions used in the paper.


\section{Bosonic string action and choice of background fields}
\setcounter{equation}{0}

The action of the closed bosonic string in the
presence of the space-time metric $G_{\mu\nu}(x)$, Kalb-Ramond
antisymmetric field $B_{\mu\nu}(x)$, and dilaton scalar field
$\Phi(x)$ is given by the following expression \cite{S}
\begin{equation}\label{eq:action2}
S = \kappa  \int_\Sigma  d^2 \xi  \sqrt{-g}  \left\{  \left[  {1
\over 2}g^{\alpha\beta}G_{\mu\nu}(x) +{\varepsilon^{\alpha\beta}
\over \sqrt{-g}}  B_{\mu\nu}(x) \right] \partial_\alpha x^\mu
\partial_\beta x^\nu +  \Phi (x) R^{(2)}  \right\} \,  ,
\end{equation}
where $\Sigma$ is the world-sheet surface 
parameterized by $\xi^\alpha=(\tau\, ,\sigma)$ [$(\alpha=0\, ,1)$,
$\sigma\in(0\, ,\pi)$], while the $D$-dimensional space-time is
spanned by the coordinates $x^\mu$ ($\mu=0,1,2,\dots,D-1$). We
denote intrinsic world sheet metric with $g_{\alpha\beta}$, and the corresponding scalar curvature with
$R^{(2)}$.

In order to keep conformal symmetry on the quantum level background fields must obey space-time field equations \cite{CFMP}
\begin{equation}\label{eq:betaG}
\beta^G_{\mu \nu} \equiv  R_{\mu \nu} - \frac{1}{4} B_{\mu \rho
\sigma} B_{\nu}{}^{\rho \sigma} +2 D_\mu a_\nu =0   \,  ,
\end{equation}
\begin{equation}\label{eq:betaB}
\beta^B_{\mu \nu} \equiv  D_\rho B^\rho{}_{\mu \nu} -2 a_\rho
 B^\rho{}_{\mu \nu} =0  \,  ,
\end{equation}
\begin{equation}\label{eq:betaFi}
\beta^\Phi \equiv 2\pi \kappa{D-26 \over 6}-R - \frac{1}{24} B_{\mu \rho \sigma}
B^{\mu \rho \sigma} -  D_\mu a^\mu + 4 a^2 =c \,  ,
\end{equation}
where $c$ is an arbitrary constant. The function $\beta^\Phi$ could be a constant because of the relation
\begin{equation}
D^\nu \beta_{\nu\mu}^G+\partial_\mu \beta^\Phi=0\, .
\end{equation}
Further, $R_{\mu \nu}$ and $D_\mu$ are Ricci tensor
and covariant derivative with respect to the space-time metric $G_{\mu\nu}$, while
\begin{equation}
B_{\mu \nu \rho}=\partial_\mu B_{\nu\rho}+\partial_\nu
B_{\rho\mu}+\partial_\rho B_{\mu\nu}\, ,\quad a_\mu=\partial_\mu
\Phi\, ,
\end{equation}
are field strength for Kalb-Ramond field $B_{\mu\nu}$ and dilaton gradient, respectively. Trivial solution of these equations is that all three background fields are constant. This case was pretty exploited 
in the analysis of the open string noncommutativity.

The less trivial case would be a case where some background fields are coordinate dependent. If we choose Kalb-Ramond field to be linearly coordinate dependent and dilaton field to be constant
then the first equation (\ref{eq:betaG}) becomes
\begin{equation}
R_{\mu \nu} - \frac{1}{4} B_{\mu \rho
\sigma} B_{\nu}{}^{\rho \sigma}=0\, .
\end{equation}
The field strength $B_{\mu\nu\rho}$ is constant and, if we assume that it is infinitesimal, then we can take $G_{\mu\nu}$ to be constant in approximation linear in $B_{\mu\nu\rho}$. Consequently, 
all three space-time field equations are satisfied. Especially, the third one is of the form
\begin{equation}
2\pi \kappa{D-26 \over 6}=c\, ,
\end{equation}
which enables us to work in arbitrary number of space-time dimensions. 

In this article we will work in $D=3$ dimensions with the following choice of background fields
\begin{equation}
G_{\mu\nu}=\left(
\begin{array}{ccc}
R_1^2 & 0 & 0\\
0 & R_2^2 & 0\\
0 & 0 & R_3^2
\end{array}\right)\, ,\quad B_{\mu\nu}=\left(
\begin{array}{ccc}
0 & Hz &0\\
-Hz & 0 &0\\
0 & 0 & 0
\end{array}\right)\, ,
\end{equation}
where $R_\mu (\mu=1,2,3)$ are radii of the compact dimensions. This choice of background fields is known in geometry as torus with flux (field strength) $H$ \cite{ALLP}. Our choice of infinitesimal $H$
can be understood in terms of the radii as that
\begin{equation}
(\frac{H}{R_1 R_2 R_3})^2=0\, .
\end{equation}
This approximation is known in literature as the approximation of diluted flux. Physically, this means that we work with the torus which is sufficiently large. Consequently, we can rescale the coordinates
\begin{equation}
x^\mu\longmapsto \frac{x^\mu}{R_\mu}\, ,
\end{equation}
which simplifies the form of the metric
\begin{equation}
G_{\mu\nu}=\left(
\begin{array}{ccc}
1 & 0 & 0\\
0 & 1 & 0\\
0 & 0 & 1
\end{array}\right)\, .
\end{equation}

The final form of the closed bosonic string action is
\begin{eqnarray}\label{eq:dejstvo}
S&=&\kappa\int_\Sigma d^2\xi \partial_+ x^\mu \Pi_{+\mu\nu}\partial_- x^\nu\\ &=&\kappa\int_\Sigma d^2\xi \left[\frac{1}{2}\left(\partial_+ x \partial_- x+\partial_+ y \partial_-y +\partial_+z \partial_-z\right)+\partial_+x Hz \partial_-y-\partial_+y Hz \partial_-x\right]\nonumber\, ,
\end{eqnarray}
where $\partial_\pm=\partial_\tau\pm \partial_\sigma$ is world-sheet derivative with respect to the light-cone coordinates $\xi^\pm=\frac{1}{2}(\tau\pm\sigma)$, $\Pi_{\pm \mu\nu}=B_{\mu\nu}\pm \frac{1}{2}G_{\mu\nu}$ and 
\begin{equation}
x^\mu=\left(
\begin{array}{c}
x\\y\\z
\end{array}\right)\, .
\end{equation}
Let us note that we do not write dilaton term because its T-dualization is performed separately within quantum formalism and here will be skipped.

\section{T-dualization of the bosonic closed string action}
\setcounter{equation}{0}

In this section we will perform T-dualization along three directions, one direction at time. Our goal is to find the relations connecting initial variables with T-dual ones called transformation laws.
Using transformation laws we will find noncommutativity and nonassociativity relations.

\subsection{T-dualization along $x$ direction - from torus with $H$ flux to the twisted torus}

Let us perform standard Buscher T-dualization \cite{B} of action (\ref{eq:dejstvo}) along $x$ direction. Note that $x$ direction is an isometry direction which means that action has a global shift symmetry, $x\longrightarrow x+a$.
In order to perform Buscher procedure, we have to localize this symmetry introducing covariant world-sheet derivatives instead of the ordinary ones
\begin{equation}
\partial_\pm x \longrightarrow D_\pm x=\partial_\pm x+v_\pm\, ,
\end{equation}
where $v_\pm$ are gauge fields which transform as $\delta v_\pm=-\partial_\pm a$. Because T-dual action must have the same number of degrees of freedom as initial one, we have to make these fields $v_\pm$
be unphysical degrees of freedom. This is accomplished by adding following term to the action
\begin{equation}
S_{add}=\frac{\kappa}{2}\int_\Sigma d^2\xi y_1 (\partial_+ v_--\partial_- v_+)\, , 
\end{equation}
where $y_1$ is a Lagrange multiplier. After gauge fixing, $x=const.$, the action gets the form
\begin{eqnarray}\label{eq:gfixac}
S_{fix}&=&\kappa\int d^2\xi \left[\frac{1}{2}\left(v_+ v_-+\partial_+ y \partial_-y +\partial_+z \partial_-z\right)+v_+ Hz \partial_-y-\partial_+y Hz v_-\right.\nonumber \\
&+&\left.\frac{1}{2}y_1 (\partial_+ v_--\partial_- v_+)\right]\, .
\end{eqnarray}

From the equations of motion for $y_1$ we obtain that field strength for the gauge field $v_\pm$ is equal to zero
\begin{equation}\label{eq:}
F_{+-}=\partial_+ v_--\partial_- v_+=0\, ,
\end{equation}
which gives us the solution for gauge field
\begin{equation}\label{eq:vpm}
v_\pm=\partial_\pm x \, .
\end{equation}
Inserting this solution for gauge field into gauge fixed action (\ref{eq:gfixac}) we obtain initial action given by Eq.(\ref{eq:dejstvo}).
Equations of motion for $v_\pm$ will lead to the T-dual action. Varying the gauge fixed action (\ref{eq:gfixac}) with respect to the gauge field $v_+$ we get
\begin{equation}\label{eq:v-}
v_-=-\partial_- y_1-2Hz \partial_- y\, ,
\end{equation}
while on the equation of motion for $v_-$ it holds
\begin{equation}\label{eq:v+}
v_+=\partial_+ y_1+2Hz \partial_+ y\, .
\end{equation}
Inserting relations (\ref{eq:v-}) and (\ref{eq:v+}) into expression for gauge fixed action (\ref{eq:gfixac}), keeping terms linear in $H$, we obtain the T-dual action
\begin{equation}\label{eq:Txdejstvo}
{}_x S=\kappa \int_\Sigma d^2 \xi \partial_+ ({}_x X)^\mu {}_x \Pi_{+\mu\nu} \partial_- ({}_x X)^\nu\, ,
\end{equation}
where subscript ${}_x$ denotes quantity obtained after T-dualization along $x$ direction and 
\begin{equation}
{}_x X^\mu=\left(
\begin{array}{c}
y_1\\ y\\z
\end{array}\right)\, .
\end{equation}
Further we have the T-dual background fields
\begin{equation}\label{eq:twisted}
{}_x\Pi_{+\mu\nu}={}_x B_{\mu\nu}+\frac{1}{2}{}_x G_{\mu\nu}\, ,\quad {}_x B_{\mu\nu}=0\, ,\quad {}_x G_{\mu\nu}=\left(
\begin{array}{ccc}
1 & 2Hz & 0\\
2Hz & 1 & 0\\
0 & 0 & 1
\end{array}\right)\, .
\end{equation}
Obtained background fields (\ref{eq:twisted}) define that what is known in literature as {\it twisted torus geometry}. String theory after one T-dualization is geometrically well defined globally and locally or, simply,
theory is geometrical (flux $H$ takes the role of connection). 

Combining the solutions of equations of motion for Lagrange multiplier  (\ref{eq:vpm}) and for gauge fields, (\ref{eq:v-}) and (\ref{eq:v+}), we get the transformation laws connecting initial, $x^\mu$, and 
T-dual, ${}_x X^\mu$, coordinates
\begin{equation}\label{eq:tlx}
\partial_\pm x \cong \pm\partial_\pm y_1\pm 2Hz \partial_\pm y\, ,
\end{equation}
where $\cong$ denotes T-duality relation.
The momentum $\pi_x$ is canonically conjugated to the initial coordinate $x$. Using the initial action (\ref{eq:dejstvo}) we get
\begin{equation}
\pi_x=\frac{\delta S}{\delta \dot x}=\kappa (\dot x-2Hz y')\, ,
\end{equation}
where $\dot A\equiv \partial_\tau A$ and $A'\equiv \partial_\sigma A$. From transformation law (\ref{eq:tlx}) it is straightforward to obtain
\begin{equation}
\dot x\cong y_1'+2Hz y'\, ,
\end{equation}
which, inserted in the expression for momentum $\pi_x$, gives transformation law in canonical form
\begin{equation}\label{eq:pix}
\pi_x\cong \kappa y_1'\, .
\end{equation}

\subsection{From twisted torus to non-geometrical $Q$ flux}

In this subsection we will continue the T-dualization of action (\ref{eq:Txdejstvo}) along $y$ direction. After $x$ and $y$ T-dualization we obtain the structure which has 
local geometrical interpretation but global omissions. Such structure is known in literature as non-geometry.

We repeat the procedure from the previous subsection and form the gauge fixed action
\begin{eqnarray}\label{eq:gfactionxy}
S_{fix}&=&\kappa\int_\Sigma d^2\xi \left[\frac{1}{2}\left(\partial_+ y_1 \partial_-y_1+v_+ v_- +\partial_+z \partial_-z\right)+\partial_+ y_1 Hz v_-+v_+ Hz \partial_- y_1\right.\nonumber \\ &+&\left.\frac{1}{2}y_2 (\partial_+ v_--\partial_- v_+)\right]\, .
\end{eqnarray}

From the equation of motion for Lagrange multiplier $y_2$
\begin{equation}
\partial_+ v_--\partial_- v_+=0 \longrightarrow v_\pm =\partial_\pm y\, ,
\end{equation}
gauge fixed action becomes initial one (\ref{eq:Txdejstvo}). Varying the gauge fixed action (\ref{eq:gfactionxy}) with respect to the gauge fields we get
\begin{equation}
v_\pm=\pm \partial_\pm y_2-2Hz \partial_\pm y_1\, .
\end{equation}
Inserting these expressions for gauge fields into gauge fixed action, keeping the terms linear in $H$, gauge fixed action is driven into T-dual action
\begin{equation}\label{eq:actionxy}
{}_{xy} S=\kappa \int d^2\xi \partial_+ ({}_{xy}X)^\mu {}_{xy}\Pi_{+\mu\nu} \partial_- ({}_{xy} X)^\nu\, ,  
\end{equation}
where
\begin{equation}
({}_{xy}X)^\mu=\left(
\begin{array}{c}
y_1\\y_2\\z
\end{array}\right)\, ,\quad {}_{xy}\Pi_{+\mu\nu}={}_{xy}B_{\mu\nu}+\frac{1}{2}{}_{xy}G_{\mu\nu}=\left(
\begin{array}{ccc}
\frac{1}{2} & -Hz & 0\\
Hz & \frac{1}{2} & 0\\
0 & 0 & \frac{1}{2}
\end{array}\right)
\, .
\end{equation}
Explicit expressions for background fields are
\begin{equation}\label{eq:bgxy}
{}_{xy} B_{\mu\nu}=\left(
\begin{array}{ccc}
0 & -Hz & 0\\
Hz & 0 & 0\\
0 & 0 & 0
\end{array}\right)=-B_{\mu\nu}\, ,\quad {}_{xy}G_{\mu\nu}=\left(
\begin{array}{ccc}
1 & 0 & 0\\
0 & 1 & 0\\
0 & 0 & 1
\end{array}\right)\, .
\end{equation}
Let us note that background fields obtained after two T-dualizations are similar to the geometric background of torus with $H$ flux, but they should be considered only locally. Their global
properties are non-trivial and because of that the term "non-geometry" is introduced.

Combining the equations of motion for Lagrange multiplier $y_2$ and for gauge fields $v_\pm$, we obtain T-dual transformation laws
\begin{equation}\label{eq:tdlaw2}
\partial_\pm y\cong \pm \partial_\pm y_2-2Hz \partial_\pm y_1\, .
\end{equation}
The $y$ component of the initial canonical momentum $\pi_y$ is a variation of the initial action with respect to the $\dot y$
\begin{equation}
\pi_y=\frac{\delta S}{\delta \dot y}=\kappa(\dot y+2 Hz x')\, .
\end{equation}
Using T-dual transformation laws (\ref{eq:tdlaw2}) we easily get
\begin{equation}
\dot y \cong y_2'-2Hz \dot y_1\, ,
\end{equation}
while from the transformation law (\ref{eq:tlx}), at zeroth order in $H$, it holds $x'\cong \dot y_1$. Inserting last two expression into $\pi_y$ we obtain transformation law in canonical form
\begin{equation}\label{eq:piy}
\pi_y\cong \kappa y_2'\, .
\end{equation}
After two T-dualizations along isometry directions, in the approximation of the diluted flux (keeping just terms linear in $H$), according to the canonical forms of the transformation laws (\ref{eq:pix}) and (\ref{eq:piy}), we see that
T-dual coordinates $y_1$ and $y_2$ are still commutative. This is a consequence of the simple fact that variables of the initial theory, which is geometrical one, satisfy standard Poisson algebra
\begin{equation}\label{eq:algebraxpi}
\left\{ x^\mu(\sigma),\pi_\nu(\bar\sigma) \right\}=\delta^\mu{}_\nu \delta(\sigma-\bar\sigma)\, ,\quad \left\{x^\mu,x^\nu\right\}=\left\{\pi_\mu,\pi_\nu\right\}=0\, ,
\end{equation}
where
\begin{equation}
\pi_\mu=\left(
\begin{array}{c}
\pi_x\\\pi_y\\\pi_z
\end{array}\right)\, .
\end{equation}

\subsection{From $Q$ to $R$ flux - T-dualization along $z$ coordinate}

In this subsection we will finalize the process of T-dualization dualizing along remaining $z$ direction. For this purpose we will use generalized T-dualization procedure \cite{DS1,DNS2,DNS3}. The result is a
theory which is not well defined even locally and is known in literature as theory with $R$-flux. 

We start with the action obtained after T-dualizations along $x$ and $y$ directions (\ref{eq:actionxy}). The Kalb-Ramond field (\ref{eq:bgxy}) depends on $z$ and it seems that it is not possible
to perform T-dualization. Let us assume that Kalb-Ramond field linearly depends on all coordinates, $B_{\mu\nu}=b_{\mu\nu}+\frac{1}{3}B_{\mu\nu\rho}x^\rho$ and check if some global transformation
can be treated as isometry one. We start with global shift transformation
\begin{equation}
\delta x^\mu=\lambda^\mu\, ,
\end{equation}
and make a variation of action
\begin{equation}
\delta S=\frac{\kappa}{3}B_{\mu\nu\rho}\lambda^\rho\int_{\Sigma}d^2\xi\partial_+x^\mu\partial_-x^\nu=\frac{2k}{3}B_{\mu\nu\rho}\lambda^\rho\epsilon^{\alpha\beta}\int_{\Sigma}d^2\xi[\partial_\alpha(x^\mu\partial_\beta x^\nu)-x^\mu(\partial_\alpha\partial_\beta x^\nu)]\, . 
\end{equation}
The second term vanishes explicitly, while the first term is surface one. Consequently, in the case of constant metric and linearly dependent Kalb-Ramond field, global shift transformation is
an isometry transformation. This means that we can make T-dualization along $z$ coordinate using generalized T-dualization procedure.

The generalized T-dualization procedure is presented in detail in Ref.\cite{DS1}. In order to localize shift symmetry of the action (\ref{eq:actionxy}) along $z$ direction we introduce covariant derivative
\begin{equation}
\partial_\pm z \longrightarrow D_\pm z=\partial_\pm z+v_\pm\, ,
\end{equation}
which is a part of the standard Buscher procedure. The novelty is introduction of the invariant coordinate as line integral
\begin{equation}
z^{inv}=\int_P d\xi^\alpha D_{\alpha}z=\int_P d\xi^+ D_+ z+\int_P d\xi^- D_- z= z(\xi)-z(\xi_0)+\Delta V\, ,
\end{equation}
where
\begin{equation}
\Delta V=\int_{P}d\xi^\alpha v_\alpha=\int_P (d\xi^+ v_++d\xi^- v_-)\, .
\end{equation}
Here $\xi$ and $\xi_0$ are the current and initial point of the world-sheet line $P$. At the end, as in the standard Buscher procedure, in order to make $v_\pm$ to be unphysical degrees of freedom
we add to the action term with Lagrange multiplier
\begin{equation}
S_{add}=\frac{\kappa}{2}\int_\Sigma d^2\xi\; y_3 (\partial_+ v_--\partial_+ v_-)\, .
\end{equation}
The final form of the action is
\begin{eqnarray}
\bar{S}&=&\kappa\int_{\Sigma}d^2\xi\left[-Hz^{inv}(\partial_+y_1\partial_-y_2-\partial_+y_2\partial_-y_1)
+\frac{1}{2}(\partial_+y_1\partial_-y_1+\partial_+y_2\partial_-y_2+D_+ zD_-z) \right. \nonumber \\
&+&\left.\frac{1}{2}\; y_3 (\partial_+ v_--\partial_- v_+)\right]\, .
\end{eqnarray}
Because of existing shift symmetry we fix the gauge, $z(\xi)=z(\xi_0)$, and then the gauge fixed action takes the form
\begin{eqnarray}\label{eq:gfaxyz}
{S}_{fix}&=&\kappa\int_{\Sigma}d^2\xi\left[-H\Delta V
(\partial_+y_1\partial_-y_2-\partial_+y_2\partial_-y_1)
+\frac{1}{2}(\partial_+y_1\partial_-y_1+\partial_+y_2\partial_-y_2+v_+v_-)\right. \nonumber\\ 
&+&\left.\frac{1}{2}y_3 (\partial_+ v_--\partial_- v_+)\right]\, .
\end{eqnarray} 
From the equation of motion for Lagrange multiplier $y_3$ we obtain 
\begin{equation}\label{eq:eomv3}
\partial_+ v_--\partial_- v_+=0\Longrightarrow v_\pm=\partial_\pm z\, ,\quad \Delta V=\Delta z\, ,
\end{equation}
which drives back the gauge fixed action to the initial action (\ref{eq:actionxy}). Varying the gauge fixed action (\ref{eq:gfaxyz}) with respect to the gauge fields $v_\pm$ we get the following equations 
of motion
\begin{equation}\label{eq:vpm3}
v_\pm = \pm \partial_\pm y_3-2\beta^{\mp}\, ,
\end{equation}
where $\beta^\pm$ functions are defined as
\begin{equation}
\beta^\pm=\pm\frac{1}{2}H(y_1\partial_\mp y_2-y_2\partial_\mp y_1)\, .
\end{equation}
The $\beta^\pm$ functions are obtained as a result of the variation of the term containing $\Delta V$
\begin{equation}
\delta_v \left(-2\kappa\int d^2\xi \varepsilon^{\alpha\beta}H\partial_\alpha y_1 \partial_\beta y_2 \Delta V\right)=\kappa\int d^2\xi \left(\beta^+ \delta v_++\beta^-\delta v_-\right)\, ,
\end{equation}
using partial integration and the fact that $\partial_\pm V=v_\pm$. Inserting the relations (\ref{eq:vpm3}) into the gauge fixed action, keeping linear terms in $H$, we obtain the T-dual action
\begin{equation}
{}_{xyz} S=\kappa \int_\Sigma d^2 \xi \partial_+ {}_{xyz}X^\mu {}_{xyz} \Pi_{+\mu\nu} \partial_- {}_{xyz} X^\nu\, ,
\end{equation}
where 
\begin{equation}
{}_{xyz}X^\mu=\left(
\begin{array}{c}
y_1 \\ y_2 \\ y_3                    
\end{array}\right)\, ,\quad {}_{xyz}\Pi_{+\mu\nu}={}_{xyz} B_{\mu\nu}+\frac{1}{2}{}_{xyz} G_{\mu\nu}\, ,
\end{equation}
\begin{equation}
{}_{xyz}B_{\mu\nu}=\left(
\begin{array}{ccc}
0 & -H\Delta \tilde y_3 & 0\\
H\Delta \tilde y_3 & 0 & 0\\
0 & 0 & 0
\end{array}\right)\, ,\quad {}_{xyz}G_{\mu\nu}=\left(
\begin{array}{ccc}
1 & 0 & 0\\
0 & 1 & 0\\
0 & 0 & 1
\end{array}\right)\, .
\end{equation}
Here we introduced double coordinate $\tilde y_3$ defined as 
\begin{equation}\label{eq:tildey}
\partial_\pm y_3\equiv \pm\partial_\pm \tilde y_3\, .
\end{equation}
Let us note that $\Delta V$ stands beside field strength $H$, which implicates that, according to the diluted flux approximation, we calculate $\Delta V$ in the zeroth order in $H$
\begin{equation}
\Delta V=\int d\xi^+ \partial_+ y_3-\int d\xi^- \partial_- y_3\, .
\end{equation}
Having this into account it is clear why we defined double coordinate $\tilde y_3$ as in Eq.(\ref{eq:tildey}). Also it is useful to note that presence of $\Delta V$, which is defined as line integral,
represents the source of non-locality of the T-dual theory. the result of the three T-dualization is a theory with $R$ flux as it is known in the literature.

Combining the equations of motion for Lagrange multiplier (\ref{eq:eomv3}), $v_\pm=\partial_\pm z$, and equations of motion for gauge fields (\ref{eq:vpm3}), we obtain the T-dual transformation law
\begin{equation}\label{eq:tlz}
\partial_\pm z\cong \pm \partial_\pm y_3-2\beta^{\mp}\, .
\end{equation}
Adding transformation laws for $\partial_\pm z$ and $\partial_-z$ we get the transformation law for $\dot z$
\begin{equation}
\dot z\cong y'_3+H(y_1 y'_2-y_2y'_1)\, ,
\end{equation}
which enables us to write down the transformation law in the canonical form
\begin{equation}\label{eq:piz}
y'_3\cong \frac{1}{\kappa}\pi_z-H(xy'-yx')\, .
\end{equation}
Here we used the expression for the canonical momentum of the initial theory (\ref{eq:dejstvo})
\begin{equation}
\pi_z=\frac{\delta S}{\delta \dot z}=\kappa \dot z\, .
\end{equation}

\section{Noncommutativity and nonassociativity using T-duality}
\setcounter{equation}{0}

In the open string case noncommutativity comes from the boundary conditions which makes that coordinates $x^\mu$ depend both on the effective coordinates and on the effective momenta \cite{gbnc}. Effective coordinates
and momenta do not commute and, consequently, coordinates $x^\mu$ do not commute. In the closed bosonic string case the logic is the same but the execution is different. Using T-duality we obtained transformation laws, (\ref{eq:tlx}), (\ref{eq:tdlaw2}) and (\ref{eq:tlz}), which
relate T-dual coordinates with the initial coordinates and their canonically conjugated momenta. In this section we will use these relations to get noncommutativity and nonassociativity relations.

\subsection{Noncommutativity relations}

Let us start with the Poisson bracket of the $\sigma$ derivatives of two arbitrary coordinates in the form
\begin{equation}\label{eq:sigmapoisson}
\{A'(\sigma),B'(\bar\sigma)\}=U'(\sigma)\delta(\sigma-\bar\sigma)+V(\sigma)\delta'(\sigma-\bar\sigma)\, ,
\end{equation}
where $\delta'(\sigma-\bar\sigma)\equiv \partial_\sigma \delta(\sigma-\bar\sigma)$. In order to find the form of the Poisson bracket
$$\{A(\sigma),B(\bar\sigma)\}\, ,$$
we have to find the form of the Poisson bracket
$$\{\Delta A(\sigma,\sigma_0),\Delta B(\bar\sigma,\bar\sigma_0)\}\, ,$$
where
\begin{equation}\label{eq:Delte}
\Delta A(\sigma,\sigma_0)=\int_{\sigma_0}^\sigma dx A'(x)=A(\sigma)-A(\sigma_0)\, ,\quad \Delta B(\bar\sigma,\bar\sigma_0)=\int_{\bar\sigma_0}^{\bar\sigma} dx B'(x)=B(\bar\sigma)-B(\bar\sigma_0)\, .
\end{equation}
Now we have
\begin{equation}
\{\Delta A(\sigma,\sigma_0),\Delta B(\bar\sigma,\bar\sigma_0)\}=\int_{\sigma_0}^\sigma dx \int_{\bar\sigma_0}^{\bar\sigma}dy\;\left[U'(x)\delta(x-y)+V(x)\delta'(x-y)\right]\, .
\end{equation}
After integration over $y$ we get
\begin{equation}
\{\Delta A(\sigma,\sigma_0),\Delta B(\bar\sigma,\bar\sigma_0)\}=\int_{\sigma_0}^\sigma dx \{U'(x)\left[\theta(x-\bar\sigma_0)-\theta(x-\bar\sigma)\right]+V(x)\left[\delta(x-\bar\sigma_0)-\delta(x-\bar\sigma)\right]\},
\end{equation}
where function $\theta(x)$ is defined as
\begin{equation}\label{eq:fdelt}
\theta(x)=\int_0^x d\eta\delta(\eta)=\frac{1}{2\pi}\left[x+2\sum_{n\ge 1}\frac{1}{n}\sin(nx)\right]=\left\{\begin{array}{ll}
0 & \textrm{if $x=0$}\\
1/2 & \textrm{if $0<x<2\pi$}\, .\\
1 & \textrm{if $x=2\pi$} \end{array}\right .
\end{equation}
Integrating over $x$ using partial integration finally we obtain
\begin{eqnarray}
&&\{\Delta A(\sigma,\sigma_0),\Delta B (\bar{\sigma},\bar{\sigma}_0)  \} = \nonumber\\
&&U(\sigma)[\theta(\sigma-\bar{\sigma}_0)-\theta(\sigma-\bar{\sigma})  ]-U(\sigma_0)[\theta(\sigma_0-\bar{\sigma}_0)-\theta(\sigma_0-\bar{\sigma})  ]\nonumber\\
&-&U(\bar{\sigma}_0)[\theta(\sigma-\bar{\sigma}_0)-\theta(\sigma_0-\bar{\sigma}_0) ]+U(\bar{\sigma})[\theta(\sigma-\bar{\sigma})-\theta(\sigma_0-\bar{\sigma})  ]\nonumber\\
&+&V(\bar{\sigma}_0)[\theta(\sigma-\bar{\sigma}_0)-\theta(\sigma_0-\bar{\sigma}_0  ]-V(\bar{\sigma})[\theta(\sigma-\bar{\sigma})-\theta(\sigma_0-\bar{\sigma})].
\end{eqnarray}
From the last expression, using the right-hand sides of the expressions in Eq.(\ref{eq:Delte}), we extract the desired Poisson bracket
\begin{equation}\label{eq:tdpoisson}
\{A(\sigma),B(\bar\sigma)\} =-[U(\sigma)-U(\bar{\sigma})+V(\bar{\sigma})  ]\theta(\sigma-\bar{\sigma})\, .
\end{equation}

Let us rewrite the canonical forms of the transformation laws, (\ref{eq:pix}), (\ref{eq:piy}) and (\ref{eq:piz}), in the following way
\begin{equation}\label{eq:yoni}
y_1'\cong \frac{1}{\kappa}\pi_x\, ,\quad y_2'\cong \frac{1}{\kappa}\pi_y\, , \quad y'_3\cong \frac{1}{\kappa}\pi_z-H(xy'-yx')\, .
\end{equation}
In order to find the Poisson brackets between T-dual coordinates $y_\mu$ we will use the algebra of the coordinates and momenta of the initial theory (\ref{eq:algebraxpi}). It is obvious that only nontrivial Poisson brackets will 
be $\{y_1(\sigma),y_3(\bar\sigma)\}$ and $\{y_2(\sigma),y_3(\bar\sigma)\}$. 

Let us first write the corresponding Poisson brackets of the sigma derivatives of T-dual coordinates $y_\mu$ using (\ref{eq:yoni})
\begin{equation}
\{y_1'(\sigma),y_3'(\bar\sigma)\}\cong \frac{2}{\kappa}Hy'(\sigma)\delta(\sigma-\bar\sigma)+\frac{1}{\kappa}Hy(\sigma)\delta'(\sigma-\bar\sigma)\, ,
\end{equation}
\begin{equation}
\{y_2'(\sigma),y_3'(\bar\sigma)\}\cong -\frac{2}{\kappa}Hx'(\sigma)\delta(\sigma-\bar\sigma)-\frac{1}{\kappa}Hx(\sigma)\delta'(\sigma-\bar\sigma)\, ,
\end{equation}
while all other Poisson brackets are zero. We see that these Poisson brackets are of the form (\ref{eq:sigmapoisson}), so, we can apply the result (\ref{eq:tdpoisson}). Consequently, we get
\begin{equation}\label{eq:nc1}
\{y_1(\sigma),y_3(\bar\sigma)\}\cong -\frac{H}{\kappa}\left[2y(\sigma)-y(\bar\sigma)\right]\theta(\sigma-\bar\sigma)\, ,
\end{equation}
\begin{equation}\label{eq:nc2}
\{y_2(\sigma),y_3(\bar\sigma)\}\cong \frac{H}{\kappa}\left[2x(\sigma)-x(\bar\sigma)\right]\theta(\sigma-\bar\sigma)\, ,
\end{equation}
where function $\theta(x)$ is defined in (\ref{eq:fdelt}). Let us note that these two Poisson brackets are zero when $\sigma=\bar\sigma$ and/or field strength $H$ is equal to zero. But if we take that
$\sigma-\bar\sigma=2\pi$ then we have $\theta(2\pi)=1$ and it follows
\begin{equation}\label{eq:nx}
\{y_1(\sigma+2\pi),y_3(\sigma)\}\cong -\frac{H}{\kappa} \left[4\pi N_y+y(\sigma)\right]\, ,
\end{equation}
\begin{equation}\label{eq:ny}
\{y_2(\sigma+2\pi),y_3(\sigma)\}\cong \frac{H}{\kappa} \left[4\pi N_x+x(\sigma)\right]\, ,
\end{equation}
where $N_x$ and $N_y$ are winding numbers defined as
\begin{equation}
x(\sigma+2\pi)-x(\sigma)=2\pi N_x\, ,\quad y(\sigma+2\pi)-y(\sigma)=2\pi N_y\, .
\end{equation}
From these relations we can see that if we choose such $\sigma$ for which $x(\sigma)=0$ and $y(\sigma)=0$ then noncommutativity relations are proportional to winding numbers. On the other side, 
for winding numbers which are equal to zero there is still noncommutativity between T-dual coordinates.

\subsection{Nonassociativity}

In order to calculate Jacobi identity of the T-dual coordinates we first have to find Poisson brackets $\{y_1(\sigma),x(\bar\sigma)\}$ as well as $\{y_2(\sigma),y(\bar\sigma)\}$. We start
with
\begin{equation}
\{\Delta y_1(\sigma,\sigma_0),x(\bar\sigma)\}=\{\int_{\sigma_0}^\sigma d\eta y'_1(\eta),x(\bar\sigma)\}\, ,
\end{equation}
and then use the T-dual transformation for $x$-direction in canonical form
\begin{equation}
\pi_x\cong \kappa y_1'\, . 
\end{equation}
From these two equations it follows
\begin{equation}
\{\Delta y_1(\sigma,\sigma_0),x(\bar\sigma)\}\cong \frac{1}{\kappa} \{\int_{\sigma_0}^\sigma d\eta \pi_x(\eta), x(\bar\sigma)\}\, ,
\end{equation}
which, using the standard Poisson algebra, produces
\begin{equation}
\{\Delta y_1(\sigma,\sigma_0),x(\bar\sigma)\}\cong -\frac{1}{\kappa}\left[\theta(\sigma-\bar\sigma)-\theta(\sigma_0-\bar\sigma)\right] \quad \Longrightarrow \{y_1(\sigma),x(\bar\sigma)\}\cong -\frac{1}{\kappa}\theta(\sigma-\bar\sigma)\, .
\end{equation}
The relation $\{y_2(\sigma),y(\bar\sigma)\}$ can be obtained in the same way. Because the transformation law for $y$-direction is of the same form as for $x$-direction, the Poisson bracket is of the same form
\begin{equation}
\{y_2(\sigma),y(\bar\sigma)\}\cong -\frac{1}{\kappa}\theta(\sigma-\bar\sigma)\, .
\end{equation}
Now we can calculate Jacobi identity using noncommutativity relations (\ref{eq:nc1}) and (\ref{eq:nc2}) and above two Poisson brackets
\begin{eqnarray}\label{eq:jacobi}
&& \{y_1(\sigma_1),y_2(\sigma_2),y_3(\sigma_3)\}\equiv \nonumber \\
&&\{y_1(\sigma_1),\{y_2(\sigma_2),y_3(\sigma_3)\}\}+\{y_2(\sigma_2),\{y_3(\sigma_3),y_1(\sigma_1)\}\}+\{y_3(\sigma_3),\{y_1(\sigma_1),y_2(\sigma_2)\}\}\cong\nonumber \\
&&-\frac{2H}{\kappa^2}\left[\theta(\sigma_1-\sigma_2)\theta(\sigma_2-\sigma_3)+\theta(\sigma_2-\sigma_1)\theta(\sigma_1-\sigma_3)+\theta(\sigma_1-\sigma_3)\theta(\sigma_3-\sigma_2)\right]\, .
\end{eqnarray}
Jacobi identity is nonzero which means that theory with R-flux is nonassociative. For $\sigma_2=\sigma_3=\sigma$ and $\sigma_1=\sigma+2\pi$ we get
\begin{equation}
 \{y_1(\sigma+2\pi),y_2(\sigma),y_3(\sigma)\}\cong \frac{2H}{\kappa^2}\, .
\end{equation}
From the last two equations, general form of Jacobi identity and Jacobi identity for special choice of $\sigma$'s, we see that presence of the coordinate dependent Kalb-Ramond field is a source of noncommutativity and nonassociativity.

\section{Conclusion}
\setcounter{equation}{0}

In this article we have considered the closed bosonic string propagating in the three-dimensional constant metric and Kalb-Ramond field with just one nonzero component $B_{xy}=Hz$. This choice of background is in accordance
with consistency conditions in the sense that all calculations were made in approximation linear in Kalb-Ramond field strength $H$. Geometrically, this settings corresponds to the torus with $H$ flux. Then we performed
standard Buscher T-dualization procedure along isometry directions, first along $x$ and then along $y$ direction. At the end we performed generalized T-dualization procedure along $z$ direction and obtained nonlocal theory with $R$ flux.
Using the relations between initial and T-dual variables, called T-dual transformation laws, in canonical form we find the noncommutativity and nonassociativity relations between T-dual coordinates.

After T-dualization along $x$ direction we obtained theory embedded in geometry known in literature as twisted torus geometry. The relation between initial and T-dual variables is trivial, $\pi_x\cong \kappa y_1'$, where $\pi_x$ is $x$ component of the canonical momentum
of the initial theory and $y_1$ is coordinate T-dual to $x$. Consequently, flux $H$ takes a role of connection, obtained theory is globally and locally well defined and commutative, because the coordinates and their canonically conjugated momenta satisfy
the standard Poisson algebra (\ref{eq:algebraxpi}).

The second T-dualization, along $y$ direction, produces nongeometrical theory, in literature known as $Q$ flux theory. The metric is the same as initial one and Kalb-Ramond field have the same form as initial up to minus sign. But, this theory
has just local geometrical interpretation. We obtained that, in approximation linear in $H$, the transformation law in canonical form is again trivial, $\pi_y\cong \kappa y_2'$, where $\pi_y$ is $y$ component od the canonical momentum of the initial theory and 
$y_2$ is coordinate T-dual to $y$. As a consequence of the standard Poisson algebra (\ref{eq:algebraxpi}), we conclude that $Q$ flux theory is still commutative. This result seems to be opposite from the result of the reference \cite{ALLP} where in detailed
calculation it is shown that $Q$ flux theory is noncommutative. The difference is in the so called boundary condition i.e. winding condition. In the Ref.\cite{ALLP} they imposed nontrivial winding condition which mixes the 
coordinates and their T-dual partners (condition given in Eq.(C.18) of Ref.\cite{ALLP}) and the result is noncommutativity. In this article the trivial winding condition is imposed on $x$ and $y$ coordinates. The consequence is that
$Q$ flux theory is commutative. But as it is written in Ref.\cite{ALLP} on page 42, "a priori other reasonings could as well be pursued".

T-dualizing along coordinate $z$ using the machinery of the generalized T-dualization procedure \cite{DS1,DNS2,DNS3} we obtain the nonlocal theory (theory with $R$ flux) and nontrivial transformation law in canonical form. Non-locality stems from the fact 
that background fields are expressed in terms of the variable $\Delta V$ which is defined as line integral. On the other side, dependence of the Kalb-Ramond field on $z$ coordinate produces the $\beta^\pm(x,y)$ functions and nontrivial transformation law for $\pi_z$.
Consequently, coordinate dependent background gives non-locality and, further, nonzero Poisson brackets of the T-dual coordinates. We can claim that there is a correlation between non-locality (R-flux theory) and closed string noncommutativity and nonassociativity.
In addition, nonzero Poisson bracket implies nonzero Jacobi identity which is a signal of nonassociativity.

From the expressions (\ref{eq:nc1}), (\ref{eq:nc2}) and (\ref{eq:jacobi}) it follows that parameters of noncommutativity and nonassociativity are proportional to the field strength $H$. That means that closed string noncommuatativity 
and nonassociativity are consequence of the fact that Kalb-Ramond field is coordinate dependent, $B_{xy}=Hz$, where $H$ is an infinitesimal parameter according to the approximation of diluted flux. Using  T-duality and trivial winding conditions
we obtained noncommutativity relations. The noncommutativity relations are zero if $\sigma=\bar\sigma$ because in noncommuatativity relations function $\theta(\sigma-\bar\sigma)$ is present, which is zero if its argument is zero.
This is also at the first glance opposite to the result of Ref.\cite{ALLP}, but, having in mind that origin of noncommutativity is not same, this difference is not surprising. If we made a round in sigma choosing 
$\sigma\to\sigma+2\pi$ and $\bar\sigma\to\sigma$, because of $\theta(2\pi)=1$, we obtained nonzero Poisson brackets. From the relations (\ref{eq:nx}) and (\ref{eq:ny}) we see that noncommutativity exists even in the case
when winding numbers are zero, noncommutativity relations still stand unlike the result in \cite{ALLP}. Consequently, we can speak about some essential noncommutativity originating from non-locality.


We showed that in {\it ordinary} space coordinate dependent background is a sufficient condition for closed string noncommutativity. Some papers \cite{minic} show that noncommutativity is possible even in the constant background case. But that could be realized 
using the {\it double space formalism}. At the zeroth order the explanation follows from the fact that transformation law in canonical form is of the form $\pi_\mu\cong \kappa y'_\mu$, where $y_\mu$ is T-dual coordinate. Forming double space
spanned by $Z^M=(x^\mu,y_\mu)$, we obtained noncommuative (double) space. In literature this kind of noncommutativity is called intrinsic one.

\appendix
\setcounter{equation}{0}

\section{Light-cone coordinates}

In the paper we often use light-cone coordinates defined as
\begin{equation}
\xi^\pm=\frac{1}{2}(\tau\pm\sigma)\, .
\end{equation}
The corresponding partial derivatives are
\begin{equation}
\partial_\pm\equiv \frac{\partial}{\partial \xi^\pm}=\partial_\tau \pm \partial_\sigma\, .
\end{equation}

Two dimensional Levi-Civita $\varepsilon^{\alpha\beta}$ is chosen in $(\tau,\sigma)$ basis as $\varepsilon^{\tau\sigma}=-1$. Consequently, in the light-cone basis the form of tensor is
\begin{equation}
\varepsilon_{lc}=\left(
\begin{array}{cc}
0 & \frac{1}{2}\\
-\frac{1}{2} & 0
\end{array}\right)\, .
\end{equation}
The flat world-sheet metric is of the form in $(\tau,\sigma)$ and light-cone basis, respectively
\begin{equation}
\eta=\left(
\begin{array}{cc}
1 & 0\\
0 & -1
\end{array}\right)\, ,\quad \eta_{lc}=\left(
\begin{array}{cc}
\frac{1}{2} & 0\\
0 & \frac{1}{2}
\end{array}\right)\, .
\end{equation}



\end{document}